\documentclass[sigconf]{acmart}

\usepackage{balance}
  
\usepackage{booktabs} 
\usepackage{subcaption}
\usepackage{cleveref}
\usepackage{multirow}
\usepackage{breqn}
\usepackage{graphicx}

\usepackage{amsmath}
\usepackage{algorithm}

\makeatletter
\def\BState{\State\hskip-\ALG@thistlm}
\makeatother

\usepackage[noend]{algpseudocode}

\def\BibTeX{{\rm B\kern-.05em{\sc i\kern-.025em b}\kern-.08emT\kern-.1667em\lower.7ex\hbox{E}\kern-.125emX}}

\copyrightyear{2019} 
\acmYear{2019} 
\setcopyright{acmlicensed}
\acmConference[SIGIR '19]{Proceedings of the 42nd International ACM SIGIR Conference on Research and Development in Information Retrieval}{July 21--25, 2019}{Paris, France}
\acmBooktitle{Proceedings of the 42nd International ACM SIGIR Conference on Research and Development in Information Retrieval (SIGIR '19), July 21--25, 2019, Paris, France}
\acmPrice{15.00}
\acmDOI{10.1145/3331184.3331372}
\acmISBN{978-1-4503-6172-9/19/07}

\crefname{figure}{Figure}{Figures}

\begin{document}
\title{Uncovering Insurance Fraud Conspiracy with Network Learning}

\author{Chen Liang}
\orcid{0000-0001-8404-1798}
\affiliation{%
  \institution{Ant Financial}
  \city{Hangzhou}
  \country{China}
}
\email{lc155190@antfin.com}

\author{Ziqi Liu}
\affiliation{%
  \institution{Ant Financial}
  \city{Hangzhou}
  \country{China}
}
\email{ziqiliu@antfin.com}

\author{Bin Liu}
\affiliation{%
  \institution{Ant Financial}
  \city{Hangzhou}
  \country{China}
}
\email{lb88701@alibaba-inc.com}

\author{Jun Zhou}
\affiliation{%
  \institution{Ant Financial}
  \city{Beijing}
  \country{China}
}
\email{jun.zhoujun@antfin.com}

\author{Xiaolong Li}
\affiliation{%
  \institution{Ant Financial}
  \city{Seattle}
  \country{USA}
}
\email{xl.li@antfin.com}

\author{Shuang Yang}
\affiliation{%
  \institution{Ant Financial}
  \city{San Francisco}
  \country{USA}
}
\email{shuang.yang@antfin.com}

\author{Yuan Qi}
\affiliation{%
  \institution{Ant Financial}
  \city{Hangzhou}
  \country{China}
}
\email{yuan.qi@antfin.com}


\begin{abstract}
Fraudulent claim detection is one of the greatest challenges the insurance industry faces. Alibaba's return-freight insurance, providing return-shipping postage compensations over product return on the e-commerce platform, receives thousands of potentially fraudulent claims everyday. Such deliberate abuse of the insurance policy could lead to heavy financial losses. In order to detect and prevent fraudulent insurance claims, we developed a novel data-driven procedure to identify groups of organized fraudsters, one of the major contributions to financial losses, by learning network information.

In this paper, we introduce a device-sharing network among claimants, followed by developing an automated solution for fraud detection based on graph learning algorithms, to separate fraudsters from regular customers and uncover groups of organized fraudsters. This solution applied at Alibaba achieves more than 80\% precision while covering 44\% more suspicious accounts compared with a previously deployed rule-based classifier after human expert investigations. Our approach can easily and effectively generalizes to other types of insurance.

\end{abstract}

%
%

\ccsdesc[500]{Social and professional topics~Financial crime}
\ccsdesc[500]{Computing methodologies~Neural networks}

\keywords{fraud detection; graph learning; network learning; insurance fraud}

\maketitle

\section{Introduction}

What if you bought a shirt but found significant color difference between the on-screen product and the real-life product? What if you discovered a less expensive alternative after purchasing a laptop? Returning the item is likely to be the first choice when shopping online. However, returning an unused item can raise lots of disputes between buyers and sellers because of the ambiguities over which party should take responsibilities. Surprisingly, most disputes focus not on whether the undamaged item should be returned, but on who should pay for return shipping costs. It takes enormous efforts and a great deal of time to resolve such disputes, especially at Alibaba\footnote{\url{https://www.alibaba.com}}, a platform with millions of sellers and diverse return policies. To resolve disputes and protect buyers' right of regret, a new form of insurance has been created. 

Return-freight insurance, designed to pay buyers the return shipping costs, has retained billions of dollars in revenue. However, the loss caused by fraudulent claims is non-trivial. Fraudsters receive shipping discounts from their partner express companies and file claims with the regular shipping price. According to the estimates of insurance experts at Alibaba, thousands of potentially fraudulent claims go undiscovered with the previous rule-based fraud detection system. Among these false claims, the most destructive ones are claimed by groups of organized fraudsters.
The need for a more powerful and more flexible fraud detection solution is significant.

\subsection{Our Fraud Detection Problem}

Fraud detection in insurance claims can be viewed as a supervised binary classification problem. We classify insurance accounts into two categories: fraudulent and regular. Labels of accounts in the training set are obtained from a formerly deployed rule-based system with some, but not sufficient, confidence. We aim to discover many more fraudulent accounts than the rule-based system while retaining high precision.

Networks provide straightforward information for describing and modeling complex relations among colluders (collaborating fraudsters). We build a device-sharing graph, a transaction graph, and a friendship graph to illustrate such relations, and apply two graph learning approaches, one based on node2vec~\cite{grover2016node2vec} and another based on GeniePath~\cite{liu2018geniepath}, to mine such information. We conduct extensive experiments to compare these approaches and describe our complete fraud detection solution which implements the device-sharing graph and our workflow deployed at Alibaba.

\subsection{Challenges in Fraud Detection}

The challenges we face that hinder the performance of fraud detection systems include \textbf{concept drift}, \textbf{label uncertainty}, and \textbf{excessive human effort}.

\textbf{Concept drift} in fraud detection refers to the phenomenon that new types of fraud evolve over time and get more and more unpredictable. It's mainly caused by the use of non-stationary features in fraud detection systems. Non-stationary behaviors, such as the number of claims made in the past month, can be easily affected when fraudsters change their tactics.
We address this problem by adding more stationary data. Relations between collaborating fraudsters are naturally stationary, e.g. device-sharing graphs.

\textbf{Label uncertainty} arises because of the usage of rule-generated labels. The formerly deployed rule-based fraud detection system outputs a risk tag for each account, say `high risk' and `no observable risk'. We are confident at `high risk' accounts, but it is unclear that whether the `no observable risk' accounts are at risk or not. In other words, the labels consist of a small amount of true
positive labels and a large amount of unknown labels. To build training labels, we randomly undersample samples from the `no observable risk' class, which is explained in the Data Preparation section.

\textbf{Excessive human effort} comes from the labeling tasks and evaluation tasks in traditional insurance fraud detection settings. As we focuses on automated risk control that with negligible human effort, our approach requires no human interventions besides a periodical evaluation (weekly or monthly) conducted by insurance professionals that samples and examines the classification results for loss estimation.

\section{Related Work}

Insurance fraud detection approaches can be generally divided into supervised learning, unsupervised learning, and a mixture of both. Popular supervised algorithms, such as Bayesian networks and decision trees have been applied or combined in \cite{viaene2004case, perez2005consolidated}. Unsupervised approaches, such as cluster analysis and outlier detection have also been applied \cite{yamanishi2004line, brockett2002fraud}. Hybrids of supervised and unsupervised algorithms have been studied, and unsupervised approaches have been used to segment insurance data into clusters for supervised approaches in \cite{brockett1998using}. 

Our two approaches fall under supervised learning and hybrids of both, respectively. Our approaches differ, as they represent data with graphs, which are one of the most natural representations of data and allow for complex analysis without simplification of data.

\section{Graph Construction}
\label{sec:graphConstr}

To address concept drift as well as to uncover organized fraudsters, 
we resort to the power of graphs that help reveal strong relations of accounts.
In this section, we construct and compare different types of graphs including a device-sharing graph, a transaction graph, and a friendship graph.
We explain which graphs fit our needs, and apply the device-sharing graph in our final fraud detection solution.


The following properties of graphs can help separate fraudulent from regular:

\begin{enumerate}
\item distance aggregation: 
closer nodes have similar labels;
\item structural differentiation: structures of organized fraudsters are different from structures of regular accounts.
\end{enumerate}

\subsection{Three Graphs}

The device-sharing graph reveals the relation of accounts sharing a device. A vertex is either a device (User Machine ID, UMID\footnote{Device fingerprints to uniquely identify devices.}) or an account. Edges only exist between a device vertex and a UMID vertex, indicating log-in activities in the history. The transaction graph shows fund exchange relations between accounts. A vertex is an account, and an edge indicates the existence of established transactions between accounts. The friendship graph is built upon friendship at Alipay, a product of Ant Financial with social networking features. We preprocess these graphs to remove singleton accounts.

\begin{table}
  \caption{Graphs for Comparison}
  \label{tab:graph-info}
  \begin{tabular}{ccccc}
    \toprule
    \textbf{Graph} & |V| & |E| & nodes & edges \\
    \midrule
    device-sharing & 3 M & 6 M & account / UMID & device usage \\ 
    transaction & 2 M & 2 M & account & fund exchange \\
    friendship & 8 M & 11 M & account & friendship \\
  \bottomrule
\end{tabular}
\end{table}

\subsection{Graph Comparison}

\begin{figure}
  \centering
  \subcaptionbox{Device-sharing: colluders\label{first-subfig}}{%
    \includegraphics[width=0.2\textwidth]{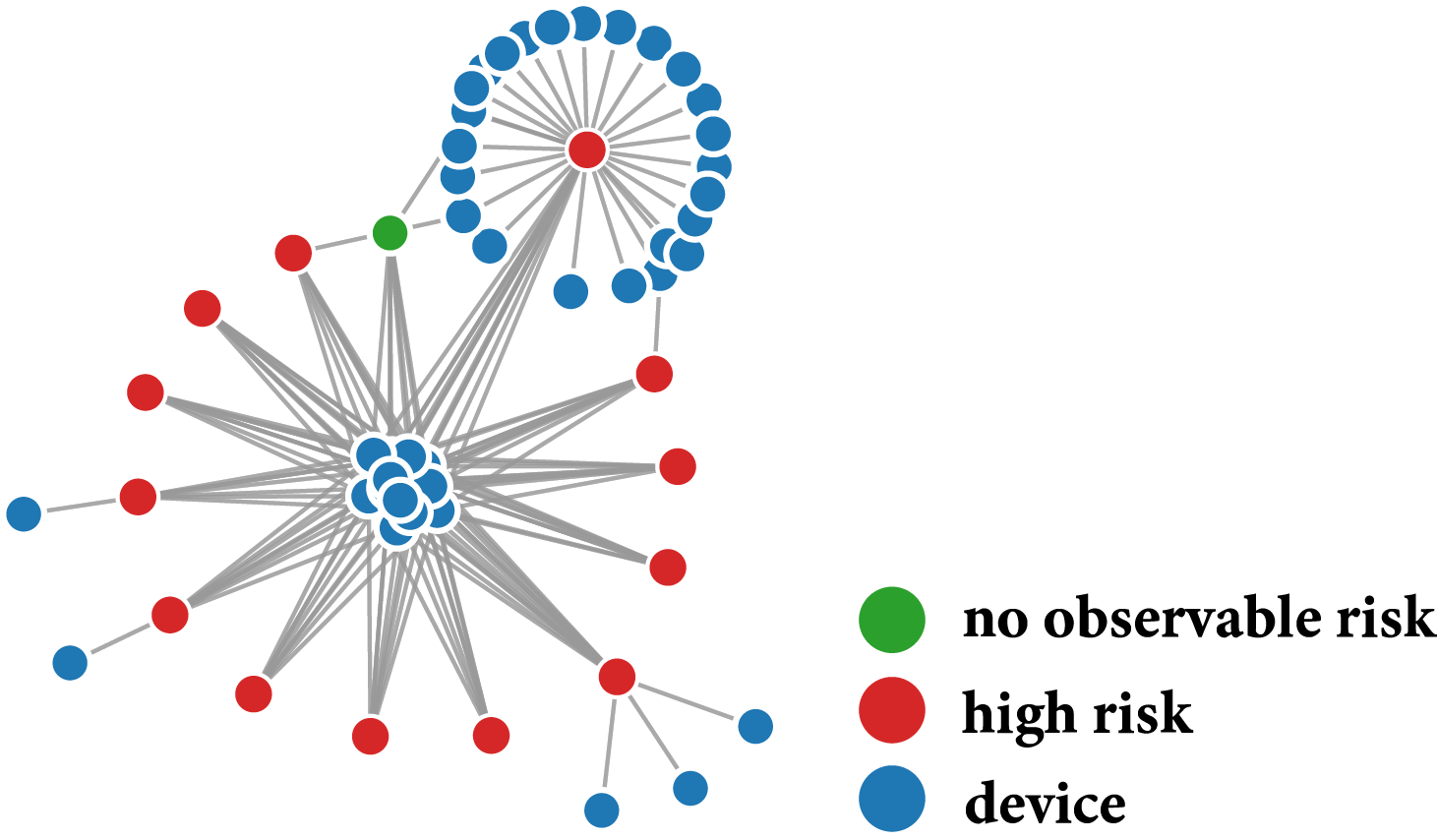}
  }
  \subcaptionbox{Device-sharing: regular\label{second-subfig}}{%
  	\makebox[0.2\textwidth][c]{\includegraphics[width=.18\textwidth]
    {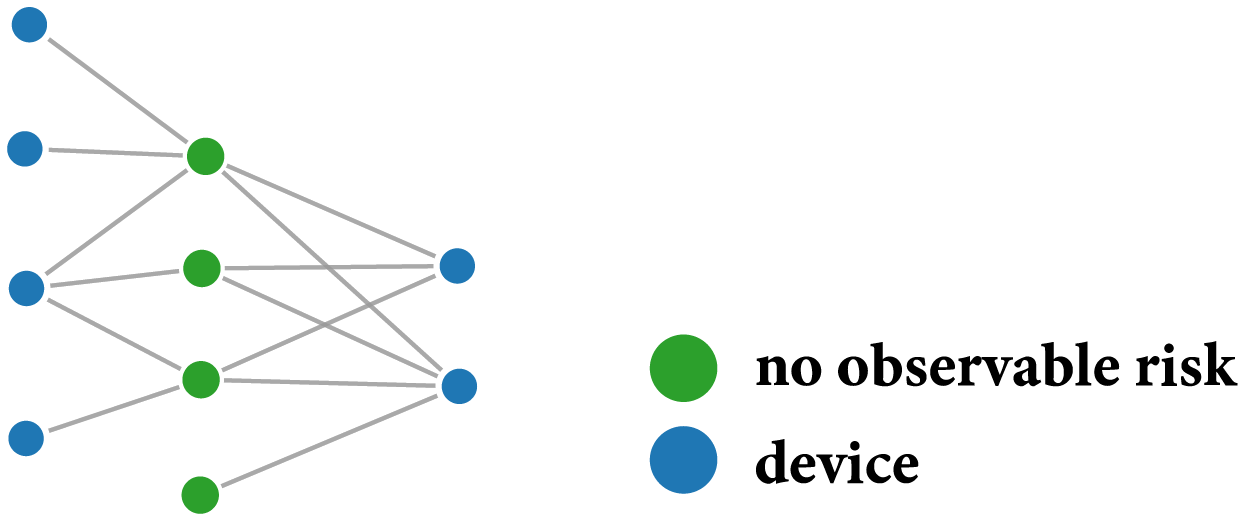}}
  }
  \subcaptionbox{Transaction: colluders\label{first-subfig}}{%
  	\makebox[0.2\textwidth][c]{\includegraphics[width=.16\textwidth]
    {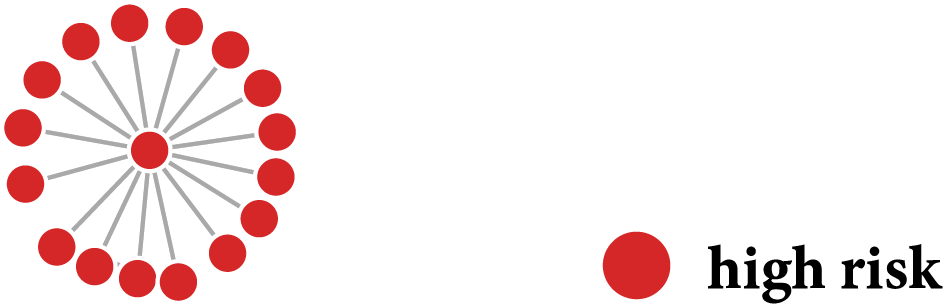}}
  }
  \subcaptionbox{Transaction: regular\label{second-subfig}}{%
    \includegraphics[width=0.2\textwidth,scale=0.4]{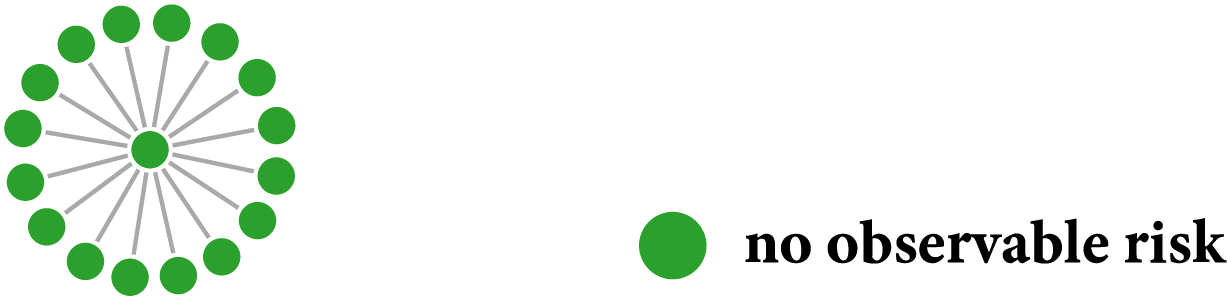}
  }
  \subcaptionbox{Friendship: colluders\label{first-subfig}}{%
    \includegraphics[width=0.2\textwidth]{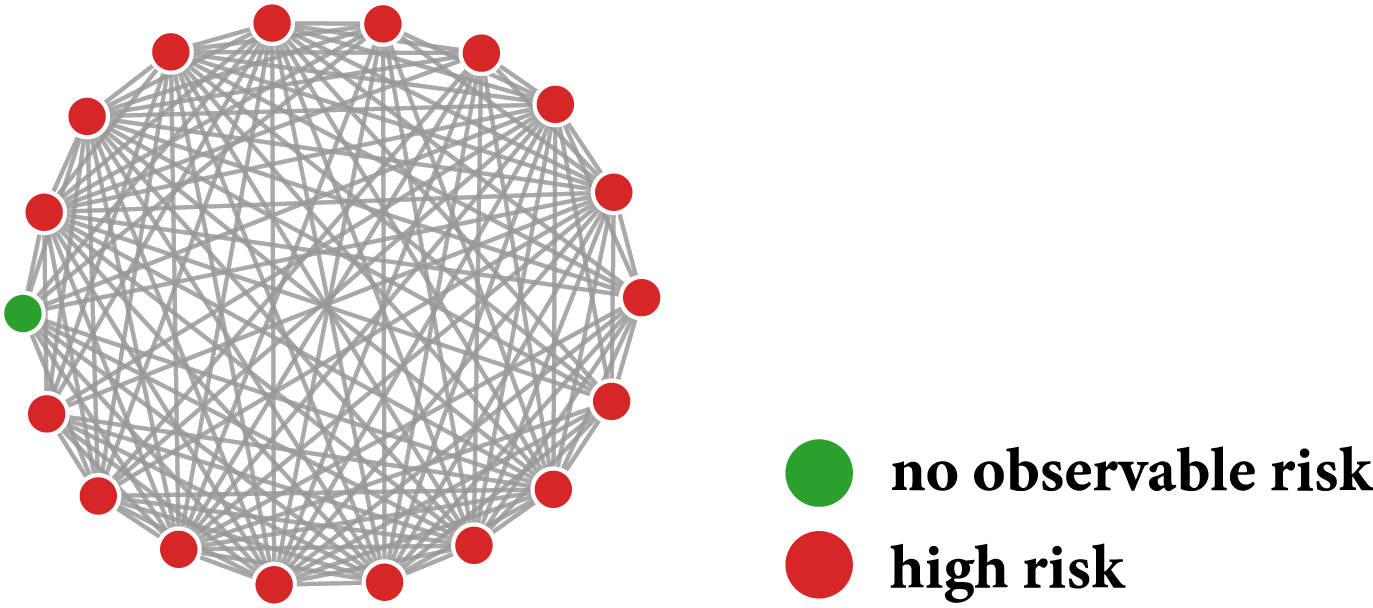}
  }
  \subcaptionbox{Friendship: regular\label{second-subfig}}{%
    \includegraphics[width=0.2\textwidth]{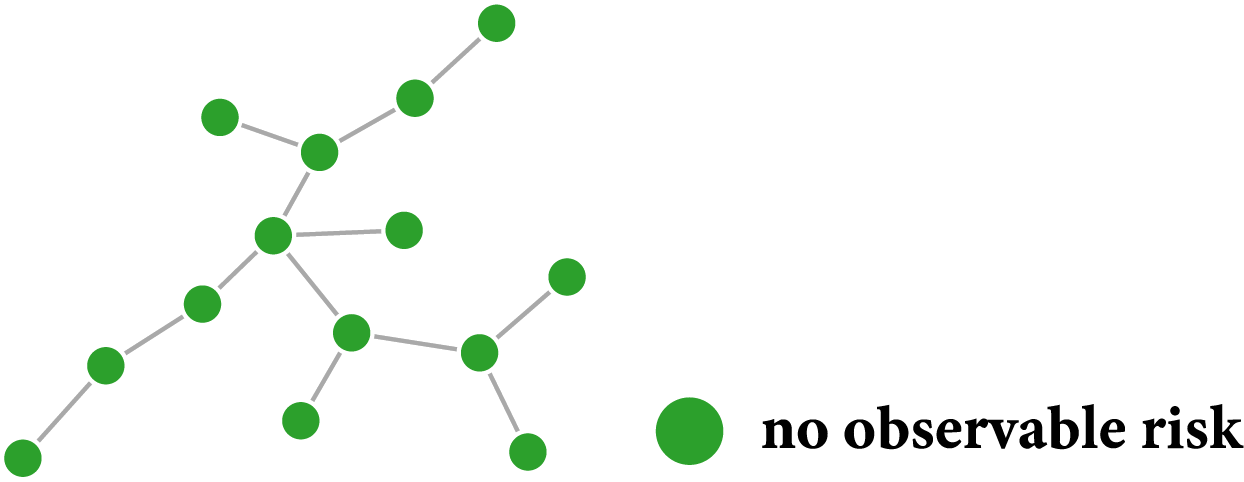}
  }
  \caption{Visualization for typical colluders and regular users in device-sharing graph, transaction graph, and friendship graph.}
  \label{fig:vis}
\end{figure}

Typical subgraphs of organized fraudsters and regular users are visualized in Figure~\ref{fig:vis}. Colluders are organized in ways that are contrasting with regular customers' as exhibited by the device-sharing graph and the friendship graph. Accounts with high risk tend to share the same group of devices and transfer funds with each other. Such patterns imply that a group of fraudsters works together to conduct frauds and split profits. The transaction graph fails to show such properties.

\begin{figure*}
\centering
  \subcaptionbox{Device-sharing\label{first-subfig-}}{%
    \includegraphics[width=0.27\textwidth]{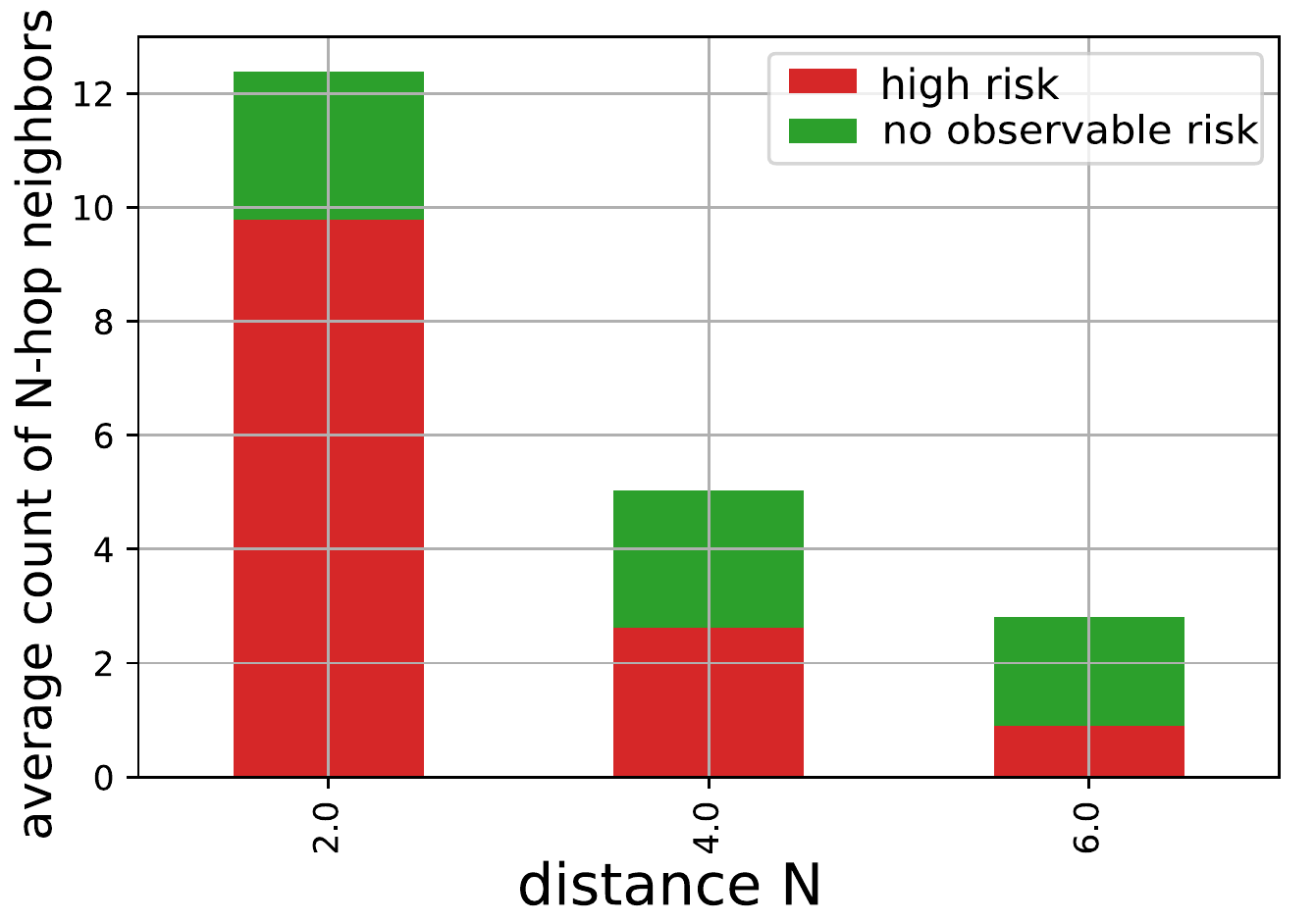}
  }
  \subcaptionbox{Transaction\label{second-subfig-}}{%
    \includegraphics[width=0.27\textwidth]{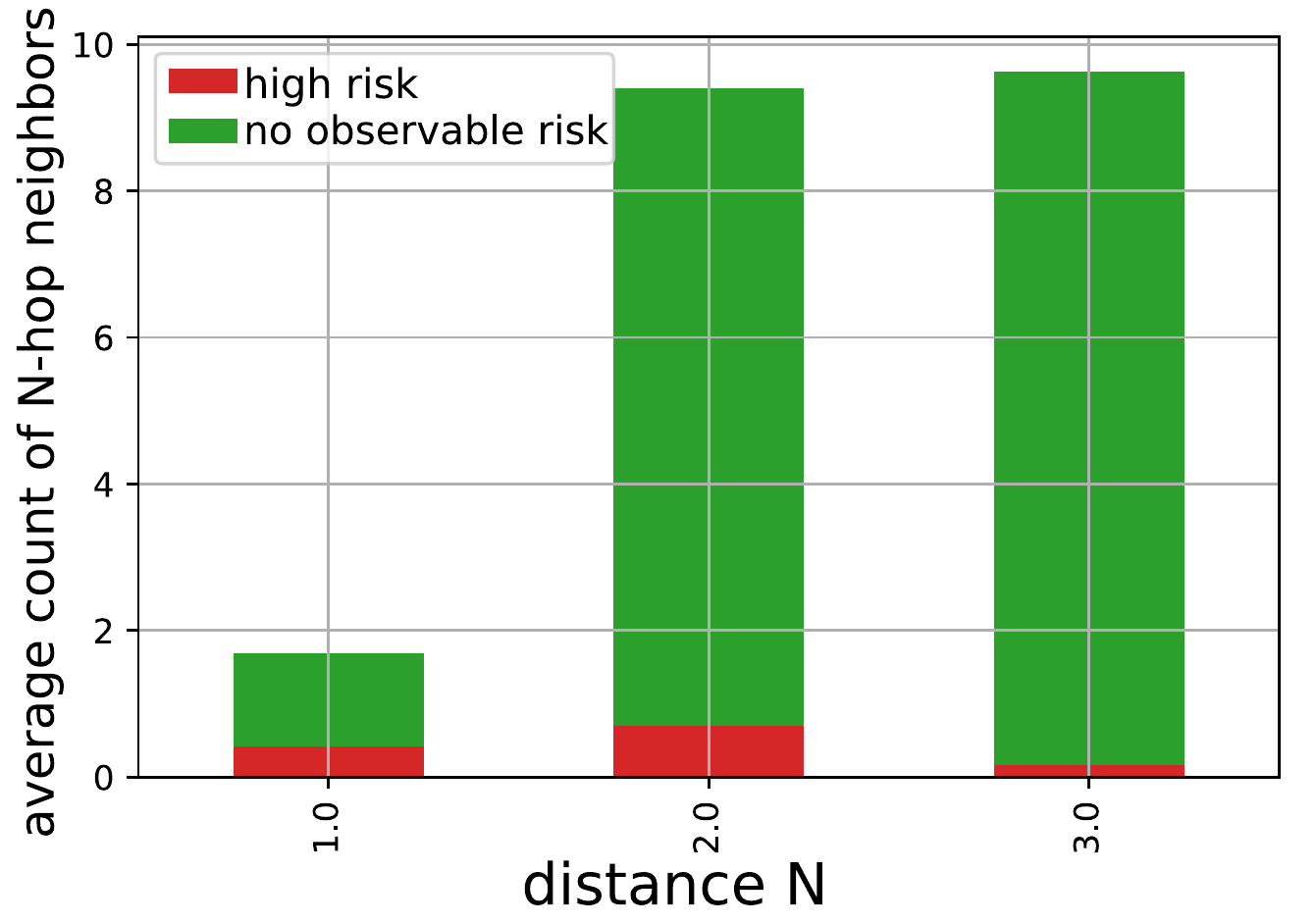}
  }
  \subcaptionbox{Friendship\label{third-subfig-}}{%
    \includegraphics[width=0.27\textwidth]{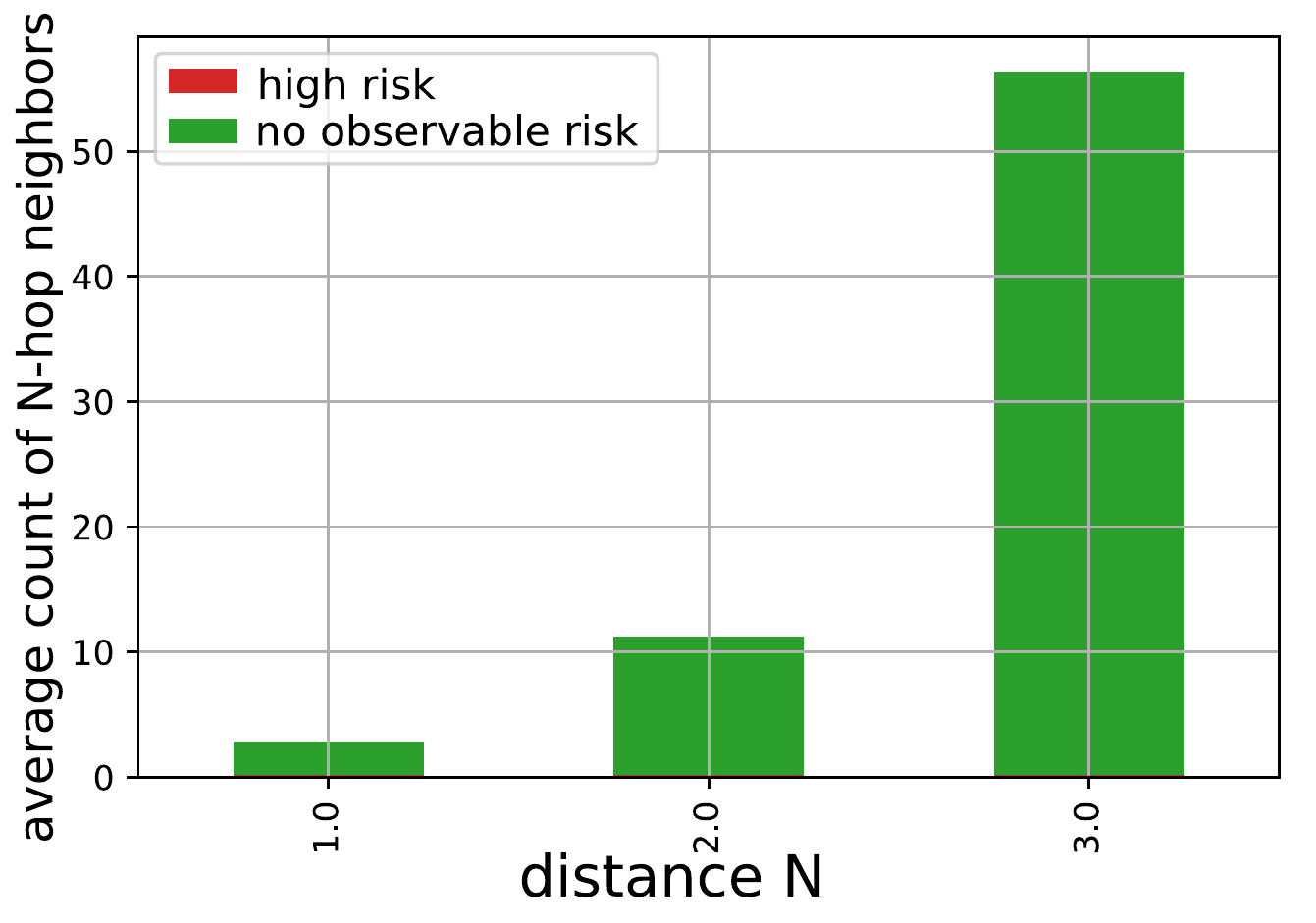}
  }
  \caption{Average number of N-hop neighbors around fraudulent accounts.}
  \label{fig:distance}
\end{figure*}

Besides, a proper graph needs to distinguish fraudulent accounts from regular accounts. Non-stationary features revealing online behavior patterns  are selected as account node features. We assume a group of fraudsters share similar behaviors. As graph neural network methods aggregate information from the neighborhood, a graph constructed with closer nodes sharing similar labels makes the classification problem easier. We measure the ability to aggregate fraudulent accounts with node distribution with respect to the distance from fraudulent nodes. The distribution is shown in Figure~\ref{fig:distance}. Fraudulent accounts gather around each other in the device-sharing graph, implying that it is more appropriate for the account classification task.

\section{Graph Learning Approach}
\label{sec:graphLearn}

Node embedding approaches and graph neural network approaches are two major techniques to understand graph information. Here we present an inductive graph neural network algorithm for the insurance fraud detection problem. A node embedding approach is briefly introduced in Section 5.2.

\subsection{Graph Neural Networks (GNNs)}

GNNs are a set of deep learning architectures that aggregate information
from nodes' neighbors using neural networks. A deeper layer in neural networks aggregates more distant neighbors, and the $t$-th layer embedding of node $u$ is 
\begin{displaymath}
  {h}_u^{(t)} = \sigma ({W}_t \cdot  \mathrm{AGG}({h}_v^{(t-1)},\forall v \in \mathcal{N}{(u)} \cup \{u\})) 
\end{displaymath} where the initial embedding ${h}_u^{(0)}={x}_u \in \mathbb{R}^{P}$ is the account feature, $h_u^{(t)} \in \mathbb{R}^K$ denotes the intermediate embedding at $t$-th layer, $\sigma$ is the activation non-linear function, and $\mathrm{AGG}(\cdot)$ is an aggregation function over neighbors that differs in GNN algorithms~\cite{liu2018geniepath}.

The GNN approach we use is based on GeniePath~\cite{liu2018geniepath}, that simply stacks adaptive path layers to aggregate each node's neighborhood based on breadth and depth exploration in the graph. For breadth exploration, it iteratively aggregates neighbors 
for $T$ times:

\begin{displaymath}
  h_u^{(t+1)} = \tanh \big( W_t \sum_{v \in \mathcal{N}(u) \cup \{u\}}{\mathrm{softmax}_{v} (\mu^\top \mathrm{tanh}({W}_s {h}_{u}^{(t)} + {W}_d {h}_{v}^{(t)})) \cdot {h}_{v}^{(t)}} \big)
\end{displaymath} This breadth-search function learns the importance of neighbors with pairwise account feature patterns. Given those hidden units $({h}_u^{(0)}, {h}_u^{(1)}, ..., {h}_u^{(T)})$ at various depths, a depth-search function $h_u = \mathrm{LSTM}({h}_u^{(0)}, {h}_u^{(1)}, ..., {h}_u^{(T)}; \phi)$ is added to further extract and filter the signals. The resulting embeddings $h_u$'s are fed to the final softmax or sigmoid layers for downstream fraud account classification tasks.

\subsection{Optimization with Label Uncertainty}
Given the final embedding $h_u$'s, we have to optimize
parameters $\theta \coloneqq \{W_t, W_s, W_d, \mu, \phi \}$. The labels used for classification are based on `risk tags' generated by a rule-based account risk indicator. We treat `high risk' accounts as fraudulent, and `no observable risk' accounts as regular. 

However, the dataset suffers from label uncertainty - the rule-based risk indicator is much more confident about `high risk' accounts being fraudulent than about `no observable risk' accounts being regular. To address this problem, `regular' accounts are sampled randomly to reduce the uncertainty of classifying a `no observable risk' account as fraudulent, as shown in the modified objective function: 

\begin{displaymath}
\begin{split}
  \mathcal{L}(\theta) & =  \min_{\theta}( \sum_{v \in \mathcal{V}_{\mathrm{fraudulent}}}\ell(\mathrm{GeniePath}({x}_v;\theta), \mathrm{fraudulent}) \\
  & + \sum_{v' \in \mathrm{sample}(\mathcal{V}_{\mathrm{regular}})}\ell(\mathrm{GeniePath}({x}_{v'};\theta), \mathrm{regular}))
\end{split}
\end{displaymath} 
Our goal is to minimize the losses caused by wrong classifications.
The chance of punishment of a false positive is controlled by the downsampling rate
in terms of the new objective function.

\section{Experiments}
\label{sec:exp}

We compare three approaches for fraud detection, two of which use graph learning algorithms, while the baseline uses account-level features only.

\subsection{Data Preparation}

Each graph we constructed contains accounts that have filed a claim within a 30-day period. Device UMIDs used by these accounts in the past 40 days are also added as graph nodes. Edges are established between account nodes and UMID nodes with login relations. Isolated subgraphs, which contain only one account node, are removed to reduce computational effort. For initial features of each account node, we collect 50 features (e.g., number of claims submitted over a month, duration as a customer, etc.), derived from insurance claim history, shipping history, and shopping history. The resulting graph contains around three million nodes and six million edges (see Table~\ref{tab:graph-info}).

\subsection{Comparison Methods}
\label{sec:methods}

We evaluate our GNN-based graph learning approach against a gradient boosted decision tree (GBDT) classifier, and against a node embedding approach. For all approaches, we calculate the probability of being at risk for each account in the test dataset, and then compare the F1 score.

\subsubsection{The GBDT approach}

The GBDT classifier uses account features as inputs without any graph structural information. 

\subsubsection{The Node Embedding Approach} 
node2vec~\cite{grover2016node2vec} assigns a low-dimensional vector to represent a graph node. It is unsupervised and only uses graph structural information. Node2vec-generated embeddings are concatenated with account features and fed to downstream classification tasks using a GBDT~\cite{zhou2017psmart}.

\subsection{Experimental Setups}

We set the same hyperparameters for all GBDT modules: 500 trees, max tree depth of 5, data sampling rate of 0.6, feature sampling rate of 0.7, and a learning rate of 0.009. We randomly sample 25\% of `no observable risk' accounts as negative samples.

\subsection{Results and Discussion}

\begin{figure}
\includegraphics[width=0.248\textwidth]{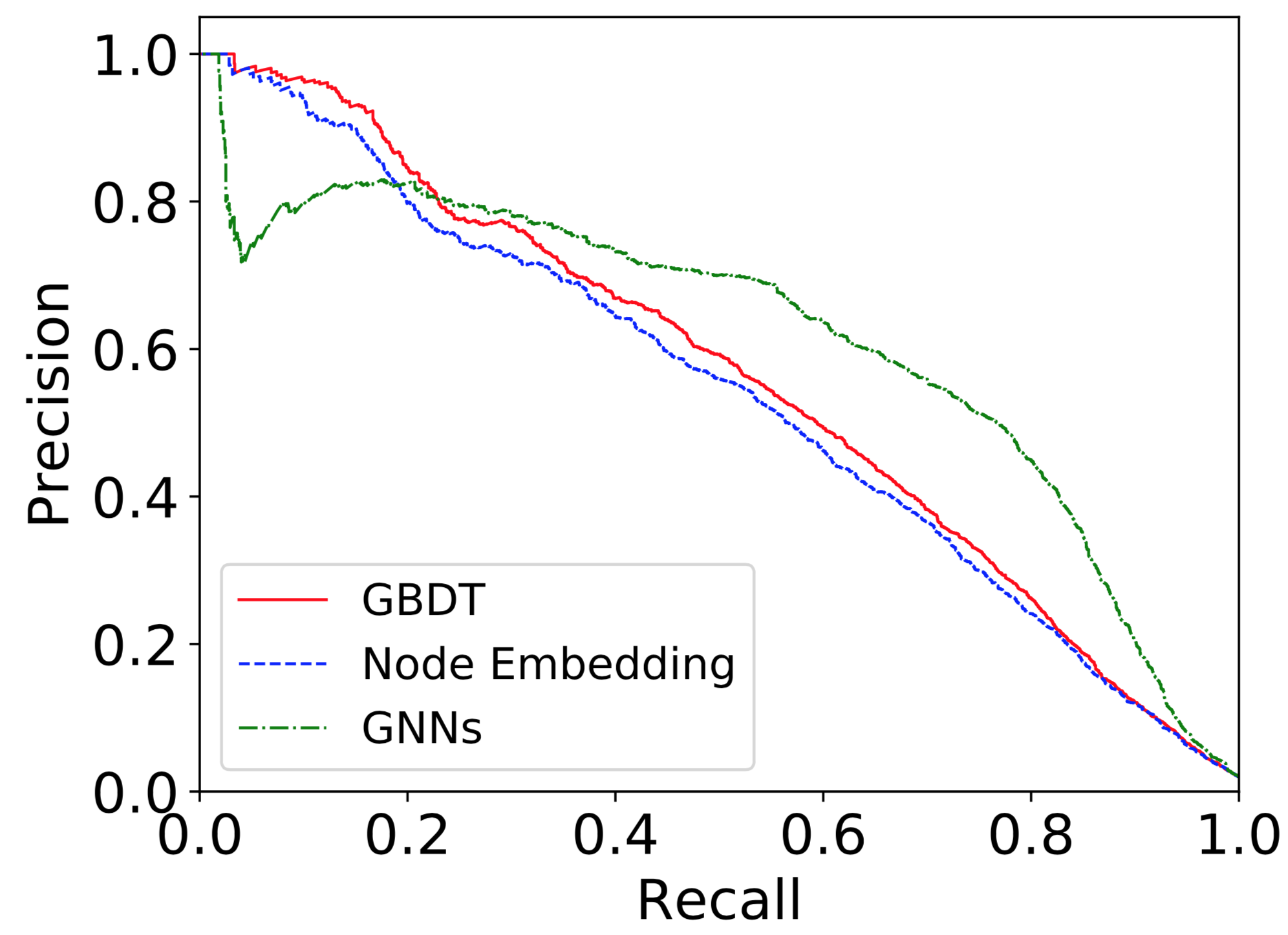}
\caption{Model comparison with the Precision-Recall curve.}
\label{fig:pr}
\end{figure}

Our results, summarized in Table~\ref{tab:results} and plotted in Figure~\ref{fig:pr}, show that the GNNs-based approach outperforms the others. Detection expansion (DE), defined as $\frac{FP + TP + FN}{TP + FN}$, indicates the ability to detect more fraudulent accounts. All of our approaches raise the coverage of fraudulent account detection by more than 40\% while GNNs-based approach has higher precision and recall at most time.

The GBDT approach is slightly better than the node embedding one. This result implies that embeddings learned solely from graph information are not as good as account features. We find out the most valuable features come from shopping history - if a user has spent a lot over the past year, we are confident he/she is not a fraudster.


\begin{table}
  \caption{Results based on Rule-based Labels.}
  \label{tab:results}
  \begin{tabular}{ccccc}
    \toprule
     &  & GBDT & Node Embedding & GNNs \\
    \midrule
    \multirow{ 2}{*}{} & F1 & 0.547 & 0.535 & 0.623 \\
    						 & DE & 1.47 & 1.44 & 1.44 \\
  \bottomrule
\end{tabular}
\end{table}

\section{Application}
\label{sec:flow}

The fraudulent claim detection system collects accounts that have filed a claim over the past months and classifies them in a batch mode that updates daily. The classification result is evaluated monthly by an insurance professional, who randomly samples and examines 300 accounts out of the reported fraudulent accounts. Daily human intervention is not necessary and human effort is enormously saved. The most recent reports show that we have achieved precision of over 80\% while covering 44\% more suspicious accounts compared with the former rule-based classifier. The estimated savings are over 10 thousand dollars per month.

\section{Conclusion} 
\label{sec:conclusion}

This paper proposes a device-sharing graph and graph learning-based approaches to address the fraud detection problem. 
It is the first paper in the literature that introduces a real-world insurance fraud detection system utilizing the strong expressiveness of graphs. Graphs have proved their power in multiple online insurance areas. 

We illustrate three types of graphs and show their advantages in separating fraudulent and regular using graph neural networks. We propose optimization algorithms for GNNs with only positive and unlabeled data. 
With proper graphs, features, and algorithms, we have achieved precision of over 80\% and covered 44\% more suspicious accounts in return-freight insurance fraud detection with automated solutions.

\bibliographystyle{ACM-Reference-Format}
\bibliography{sample-bibliography}

\end{document}